\documentclass[12pt]{article} \def\theequation{\arabic{section}.\arabic{equation}}
\newcounter{rown}

\usepackage[centertags]{amsmath}
\usepackage{cite}
\usepackage{color}
\usepackage{amsfonts}
\usepackage{amssymb}
\usepackage{amsthm}
\usepackage{newlfont}

\def\be{\begin{equation}}
\def\ee{\end{equation}}
\def\bi{\begin{itemize}}
\def\ei{\end{itemize}}
\def\ben{\begin{enumerate}}
\def\een{\end{enumerate}}
\begin{document}
\renewcommand{\theequation}{\arabic{section}.\arabic{equation}}
\title{{\bf Bosonic D=11 supergravity from a generalized Chern-Simons action}}
\vskip 6cm
\author{  D. Camarero${}^\dagger$, J.A. de Azc\'{a}rraga${}^\star$, J. M. Izquierdo${}^\dagger$,\\
${}^\dagger$ Departamento de F\'{\i}sica Te\'orica, Universidad de Valladolid, \\
47011-Valladolid, Spain\\
${}^\star$Departamento de F\'{\i}sica Te\'orica and IFIC (CSIC-UVEG), \\
 46100-Burjassot (Valencia), Spain}
\date{June 27, 2017}
\maketitle
\vskip 1cm
\begin{abstract}
It is shown that the action of the bosonic sector of $D=11$ supergravity
may be obtained by means of a suitable scaling of the originally
dimensionless fields of a generalized Chern-Simons action. This
follows from the eleven-form CS-potential of the most general linear combination
of closed, gauge invariant twelve-forms involving the $sp(32)$-valued two-form
curvatures supplemented by a three-form field. In this construction,
the role of the skewsymmetric four-index auxiliary function needed
for the first order formulation of $D=11$ supergravity is played
by the gauge field associated with the five Lorentz indices generator
of the bosonic $sp(32)$ subalgebra of $osp(1|32)$.
\end{abstract}
\newpage

\section{Introduction}
It is known \cite{AP-86,W-88,AP-89,AI12} that various $D=3$ (super)gravities
are actually Chern-Simons (CS) theories based on Lie superalgebras.
Although supergravities in $D>3$, $D$ odd, do not have a true CS nature,
it has been argued that certain CS theories may be
related to supergravities for odd $D>3$ dimensions. These CS
theories have been generically called `CS supergravities' \cite{Cha89-90,BTZ96,TZ98}
(see \cite{EZ08} for further references).

CS actions are constructed (see {\it e.g.}~\cite{CUP}) as follows. Let
$A^i$, $F^i$ ($i=1,\dots ,\textrm{dim} \, \mathcal{G}$) be the Maurer-Cartan (MC)
gauge fields and curvatures associated with a Lie algebra
$\mathcal{G}$ in a certain basis. Then, the $2\ell$-form (the exterior product symbol
$\wedge$  will be omitted throughout)
\begin{equation}
\label{Hk}
H= k_{i_1\dots 1_\ell} F^{i_1}  \dots  F^{i_\ell} \quad,
\end{equation}
where $k_{i_1\dots 1_\ell}$ are the coordinates of
a symmetric invariant tensor of order $\ell$, is closed and gauge invariant.
Since a gauge free differential algebra is contractible, $H$
is also exact, $H=dB$, and the potential $B$ defines a Chern-Simons $(2\ell-1)$-form,
which is gauge invariant up to an exterior differential. Then,
the CS action is given by the integral
\begin{equation}
\label{intB}
 I_{CS}= \int_{\mathcal{M}^{2\ell-1}} B
 \end{equation}
over a $(2\ell-1)$-dimensional manifold $\mathcal{M}^{2\ell-1}$; it
is gauge invariant up to non-trivial topological situations ignored
in this paper.

The possible connection between CS supergravity and the actual supergravities for $D>3$
suggested in refs. \cite{Horava,Nastase,IMPRS09,IR12} (see \cite{bañados} for
another connection in $D=11$ based on the comparison of the linearized models)
is best analyzed by expressing the gauge fields and curvatures
associated with the superalgebra $\mathcal{G}$
in terms of supermatrices $\mathbb{A}$ and $\mathbb{F}$, with
one- and two-form fields entries respectively. This is the case for $D=3$ and
$\mathcal{G}=osp(p|2) \oplus osp(q|2)$, for $D=5$ and $\mathcal{G}=su(1|2,2)$ and for
$D=11$ and $\mathcal{G}=osp(1|32)$ (or $osp(1|32)\oplus osp(1|32)$).
$H$ is typically of the form $H= \mathrm{Tr}(\mathbb{F}^\ell)$
where $\mathrm{Tr}$ denotes the supertrace, although other
non-primitive, closed gauge invariant forms will be considered below.
Depending on the case, the MC one-form gauge fields of the superalgebras may,
or may not, correspond to the fields of $D$-dimensional supergravities. In the second
and almost general case, the association between  `CS supergravities'
and the standard supergravities in $D$ dimensions fails. Let us show this by
summarizing the $D$=3,5 and 11 cases.

We use mostly plus metric throughout.

\subsection{The $D=3$ case.}
Let us first consider the simplest algebra
$\mathcal{G}=osp(1|2)\oplus sp(2)$  ({\it i.e.} $p=1$, $q=0$ above).
The $osp(1|2)$ and $sp(2)$ gauge fields, denoted $\mathbb{A}$ and
$\widetilde{\mathbb{A}}$ respectively, can be written in matrix form as
\begin{equation}
\label{d3matrix}
    \mathbb{A}=\left(
        \begin{array}{cc}
          f & \xi \\
          \bar{\xi} & 0 \\
        \end{array}
      \right)
      \ ,\; f=f_a\gamma^a \quad ; \quad  \widetilde{\mathbb{A}} = \widetilde{f} \ , \quad
      \widetilde{f} = \widetilde{f}_a \gamma^a \ ,
\end{equation}
where $\xi$ is a two-component Grassmann odd Majorana spinor form and $\gamma^a$ are the
$2\times 2$ $D=2$ gamma matrices. Note that $osp(1|2)$ alone would not provide enough fields for
$D=3$ supergravity and that all fields $f_a$, $\xi$ and $\widetilde{f}_a$
in \eqref{d3matrix} are necessarily dimensionless; to define
`physical' one-form fields, we introduce a scale parameter $\lambda$,
$[\lambda]= L^{-1}$. We use geometrized units for which $c = 1 = G$, so that all the
quantities have physical dimensions in terms of powers of length; with them,
the dimensions of an action in $D$-dimensional spacetime is $L^{(D-2)}$.
The new fields $\omega_a$,  $e_a$, and $\psi$  obtained from $f$, $\xi$ and
$\widetilde{f}$ are then defined by
\begin{equation}
\label{rescale3}
    f_a = \omega_a +\lambda e_a \ , \quad \widetilde{f}_a= \omega_a \ , \quad \xi
    = \lambda^{\frac{1}{2}} \psi \ ,
\end{equation}
so that they have the right dimensions $[\omega_a]=L^0$, $[e_a]= L^1$ and
$[\psi]=L^{\frac{1}{2}}$ to be identified with the fields of $D=3$, $N=1$
supergravity in the first order formulation.

The action is constructed starting from the closed, invariant polynomial four-form
\begin{equation}
\label{H3inicial}
    H(f,\widetilde{f},\xi ;\alpha)=
    \textrm{Tr}(\mathbb{F}^2) + \alpha \textrm{Tr}(\widetilde{\mathbb{F}}^2)\ ,
\end{equation}
where $\alpha$ is a dimensionless constant and
\begin{equation}\label{MF3}
   \mathbb{F}=d \mathbb{A} +  \mathbb{A}^2 = \left(
                 \begin{array}{cc}
                   df + f^2 + \xi \bar{\xi} &d\xi+ f\xi \\
                   d\bar{\xi} + \bar{\xi} f & 0 \\
                 \end{array}
               \right)
                \quad , \quad
                \widetilde{\mathbb{F}} = d\widetilde{f} + \widetilde{f}^2  \  .
\end{equation}
Inserting \eqref{rescale3} into \eqref{H3inicial}
and collecting the terms in equal powers of $\lambda$ gives
\begin{equation}
\label{expan3}
    H(\omega,e,\psi;\lambda, \alpha)= H_0 + \lambda H_1 +\lambda^2 H_2 + \lambda^3 H_3 \ ,
\end{equation}
where $H_0=H_0(\omega, \alpha)$ only since $H(\omega,e,\psi;\lambda , \alpha)$ is dimensionless
and $H_{1,2,3}\neq H_{1,2,3}(\alpha)$.
We note in passing that this re-scaling in $\lambda$ is the starting point of the
(super)Lie algebra expansions procedure, introduced in \cite{Sakaguchi} and
considered in general in \cite{AIMV}, by which new (super)algebras may be obtained from
a given one. Note that, unlike in the contraction of algebras, where
the dimensions of the original algebra and that of the contracted one are
necessarily equal, the dimension of the expanded algebra is usually higher
since the expansion process is not dimension-preserving in general\footnote{It is
terminologically unfortunate that algebras of different dimensions are sometimes
said to be related by so-called `generalized' \.In\"on\"u-Wigner contractions. There
are, of course, generalizations of the original I-W contraction procedure with
respect to a subalgebra, but these are also dimension-preserving, as it
corresponds to the mathematical idea of contraction (see e.g. \cite{AIPV07}).}
(see \cite{AIMV,AIPV07} for details).

By construction, the above two-form $H$ and the associated CS action are $osp(1|2)\oplus sp(2)$
gauge-invariant. In particular, the local supersymmetry transformations under the odd
dimensionless gauge parameter $\eta$ that corresponds to the gauge field $\xi$ are,
written in terms of $\epsilon = \lambda^{-\frac{1}{2}}\eta$, $[\epsilon]=L^{1/2}$,
\begin{eqnarray}
\label{3transf}
\nonumber
   \delta_\epsilon e^a &=& \overline{\psi} \gamma^a \epsilon\ , \\
\nonumber
   \delta_\epsilon \psi &=& D \epsilon + \lambda e_a \gamma^a \epsilon \ , \\
   \delta_\epsilon \omega^a &=& 0 \ ,
\end{eqnarray}
where $D=d-\omega_a\gamma^a$ is the Lorentz covariant derivative.
Since $\omega^a$ is supersymmetry invariant, so is $H_0$ which only contains this field.
Thus, the action obtained from $H(\omega,e,\psi;\lambda,\alpha)-H_0(\omega,\alpha)$
is invariant under the local supersymmetry transformations \eqref{3transf},
and provides the first order formulation of  $(1,0)$ $D$=3 $AdS$ supergravity.
Moreover, the leading $\lambda$ term in $H-H_0$, $H_1$, is also invariant under
the transformations \eqref{3transf} for $\lambda=0$, and hence provides the
action for $D=3$ Poincar\'e supergravity; this will not be the case for higher $D$.
Also, as noted in \cite{AP-89}, in the general $(p,q)$ case the action contains a term
that comes from $H_0$ which is not invariant under the $\epsilon$ gauge transformation
that cannot be ignored and the linear term in $\lambda$
does not yield Poincar\'e supergravity. In this case, a proper Poincar\'e
limit may still be taken by enlarging $\mathcal{G}$ as $\mathcal{G}$ to
 $osp_+(p|2) \oplus osp_-(q|2) \oplus so(p) \oplus so(q)$, and adding to $H$ the
 two invariant $so(p)$ and $so(q)$-valued four-forms \cite{HIPT}\cite{AI12}.

\subsection{The $D=5$ case.}
The next simplest case is $D=5$. The smallest real superlgebra that contains
the $AdS_5$ one $so(4,2)\sim su(2,2)$ is the 24-dimensional
$\cal{G}= su(1|2,2)$. A $su(1|2,2)$-valued form
can be written in the form
\begin{equation}
\label{su14}
   \mathbb{A}= \left(
        \begin{array}{cc}
          f & \xi \\
          i \bar{\xi} & 4if_0 \\
        \end{array}
      \right)
      \quad ,\quad f= if_0 +f_a \gamma^a + \frac{1}{4} f_{ab} \gamma^{ab}\ ; \
      \mathbb{F}= d\mathbb{A} + \mathbb{A}^2 \ ,
\end{equation}
where $\gamma^a$, $a=0,\dots, 4$ are $4\times 4$ gamma matrices, $\xi$ is a
four-component spinor form and $\bar{\xi}$ its adjoint. Let us
introduce again $\lambda$, $[\lambda]=L^{-1}$, and new fields $e_a$,
$\phi$, $\omega_{ab}$ and $\psi$, with dimensions $1$, $1$, $0$ and $1/2$
respectively, through the scalings $f_0= \lambda \phi$, $f_a=\lambda e_a$,
$f_{ab}=\omega_{ab}$, $\xi=\lambda^{\frac{1}{2}}\psi$. We now express
the 16 real bosonic fields 1($\phi$)+5($e$)+10($\omega$)
and the 4 complex fermionic ones $\psi$  (8 real) associated with
the supergroup parameters in the form
\begin{equation}
\label{resca5}
    f= i \lambda \phi +  \lambda e_a \gamma^a
    + \frac{1}{4} \omega_{ab} \gamma^{ab} \ \; , \ \;   \xi= \lambda^{\frac{1}{2}}\psi \ .
\end{equation}
Using these expressions in $\mathbb{F}$ and $H=$Tr$(\mathbb{F}^3)$  and collecting the
different powers in $\lambda$ we obtain
\begin{equation}
\label{expan5}
 H(\phi, e,\omega,\psi)= H_0 + H_1 \lambda +H_2 \lambda^2 + H_3 \lambda^3 + H_4 \lambda^4 + H_5 \lambda^5 \ ,
\end{equation}
where $H_0$=$H_0(\omega)$ and $H_i$, $=1,\dots 5$, depend on the gauge fields
$e_a$, $\phi$, $\omega_{ab}$ and $\psi$.

The term $H_3$ in $\lambda^3$ has the right dimension
$[H_3]=L^{D-2}=L^3$ for a $D$=5 action. Therefore, it makes sense
comparing the CS action obtained from $H_3$ with that of simple $D=5$
supergravity which, in the first order formulation, has the same spacetime
fields content; including also the terms proportional to $\lambda^4$ and $\lambda^5$
and retaining only the last three terms would lead (removing a common $\lambda^3$ factor)
to an action with a `cosmological constant' term in $\lambda^2$ coming from $H_5$
(as it would be similarly the case taking the higher order
terms in $D=3$ \cite{AP-86}). However, here there is no reason why
local supersymmetry should be preserved by selecting any group
of terms in \eqref{expan5}: since the $su(1|4)$ $\epsilon$ gauge
transformations in terms of the rescaled fields depend on $\lambda$,
\begin{eqnarray}
\label{5eps}
\delta_\epsilon \phi &=& - \frac{1}{4}\left( \bar{\epsilon} \psi - \bar{\psi} \epsilon \right) \nonumber \\
\delta_\epsilon e^a &=& -\frac{i}{4}\left( \bar{\epsilon}\gamma^a \psi - \bar{\psi}\gamma^a \epsilon \right)
\nonumber\\
\delta_\epsilon \omega^{ab} &=& \frac{i\lambda}{2} \left( \bar{\epsilon}\gamma^{ab} \psi - \bar{\psi}
\gamma^{ab} \epsilon \right)\nonumber\\
\delta_\epsilon \psi &=& d\epsilon + \frac{1}{4} \omega_{ab}\gamma^{ab}\epsilon + \lambda\left( -3i\phi
+ e_a\gamma^a\right)\epsilon \ ,
\end{eqnarray}
the individual terms are not invariant separately. The leading $H_0$ term will be invariant
under the above gauge algebra for $\lambda=0$, but this will
not be the case for the other terms including the one with the correct dimension $H_3$.
In fact, it is easily seen that the $H_3$ term in \eqref{expan5} does not lead to
$D=5$ supergravity. The quickest way to see it is by noticing that this  $H_3$ term coming from
the $su(1|4)$ based CS action is not gauge invariant under the one-dimensional subgroup of
transformations $\varphi$ corresponding to the field $\phi$,
$\delta_\varphi \phi = d \varphi$, in contrast with the action of the $D=5$ supergravity.

\subsection{The $D=11$ case: preliminary considerations.}
\label{prelim}
The $D$=11 AdS algebra $so(2,10)$ is contained in $sp(32)$, which is
of dimension (32+1)$\cdot$16. The relevant superalgebra in this
case would be, in principle, the smallest one that contains $sp(32)$,
namely $osp(1|32)$, of dimension 528+32=560. A convenient way of describing
its elements is provided by the $osp(1|32)$-valued one-form gauge
field supermatrix
$\mathbb{A}$ given by
\begin{equation}\label{A32}
    \mathbb{A}=\left(
        \begin{array}{cc}
          f & \xi \\
          \bar{\xi} & 0 \\
        \end{array}
      \right) \ , \quad
      f= f^a\gamma_a + \frac{1}{4} f^{ab}\gamma_{ab}
      + f^{a_1\dots a_5} \gamma_{a_1\dots a_5} \; ,
\end{equation}
where $\gamma_a$ are the $32\times 32$ gamma matrices and $\xi$ is a $32$-component
Majorana spinor one-form. Clearly, the $osp(1|32)$-valued
one-forms in \eqref{A32} cannot be identified with the one-form fields $e_a$,
$\omega_{ab}$, $\psi^\alpha$ and the {\it three}-form field $A$ of
Cremmer-Julia-Scherk (CJS) $D$=11 supergravity \cite{CJS}.

   One could think of using  two copies \cite{Horava} $osp(1|32), \widetilde{osp}(1|32)$,
to write the gauge fields $f^a$, $\tilde{f}^a$, $f^{ab}$, $\tilde{f}^{ab}$,
$f^{a_1\dots a_5}$, $\widetilde{f}^{a_1\dots a_5}$, $\xi^\alpha$, $\widetilde{\xi}^\alpha$, as linear
combinations of new fields $e^a$, $B^a$, $\omega^{ab}$, ${B}^{ab}$,
${B}^{a_1\dots a_5}$, ${B'}^{a_1\dots a_5}$,
$\psi^\alpha$, ${\psi'}^\alpha$, with dimension $L$ except for
$[\psi^\alpha]= L^{\frac{1}{2}}$, $[{\psi'}^\alpha]= L^{\frac{3}{2}}$,
$[\omega^{ab}]= L^0$ (and perhaps ${B'}^{a_1,\dots,a_5}$ and $B^a$)
using the scale factor $\lambda$, $[\lambda]=L^{-1}$.
It was conjectured in \cite{Horava} that the three-form field $A$ could be a
composite of $e^a$, ${B}^{ab}$, $B^{a_1\dots a_5}$,
$\psi^\alpha$, ${\psi'}^\alpha$ as explicitly considered in \cite{DF82}.
A closed, $osp(1|32)$ gauge invariant twelve-form $H$ has the
general expression
\begin{equation}
\label{H11}
   H= \textrm{Tr}(\mathbb{F}^6) + \alpha \textrm{Tr}(\mathbb{F}^2) \textrm{Tr}(\mathbb{F}^4) +\beta
   \left(\textrm{Tr}(\mathbb{F}^2)\right)^3 \ ,
\end{equation}
where $\mathbb{F}= d\mathbb{A}+ \mathbb{A}^2$. The corresponding form $\widetilde{H}$
for $\widetilde{osp}(1|32)$ is expressed similarly in terms of
$\widetilde{\mathbb{F}}= d\widetilde{\mathbb{A}}+\widetilde{\mathbb{A}}^2$.
Then, introducing $H'(\lambda)=H(\lambda)+\widetilde{H}(\lambda)$ and collecting the
different powers of $\lambda$ we can write
\begin{equation}
\label{lambdaexp11}
    H'(\lambda)=H(\lambda)+\widetilde{H}(\lambda) =
    H'_0 + \dots + H'_9 \lambda^9 + H'_{10} \lambda^{10} + H'_{11} \lambda^{11} \ .
\end{equation}

It was conjectured \cite{Horava} that the $H'_9$ term would depend on $\omega^{ab}$,
$e^a$ and $\psi^\alpha$, with the remaining fields either included in
$A=A(e^a,{B}^{ab},B^{a_1\dots a_5},\psi^\alpha, {\psi'}^\alpha)$
or absent, and that it  would also be invariant under
 local supersymmetry.
However, this has not been verified, and there are arguments against this being the case.
First, the bosonic and fermionic on-shell degrees of freedom do not match
unless there is a large hidden extra gauge symmetry.
To be more precise, let us consider Horava's choice of $osp(1|32)\oplus \widetilde{osp}(1|32)$
and possible gauge action depending on  $e^a$, $B^a$, $\omega^{ab}$, $B^{ab}$,
${B}^{a_1\dots a_5}$, ${B'}^{a_1\dots a_5}$,
$\psi^\alpha$ and ${\psi'}^\alpha$ with the following assumptions:
(a) the action corresponding to $H'_9$ has the gauge symmetries of the above fields
realized in the generic form $\delta A^i = d\alpha^i+\dots $;
(b) the fields $B^a$ and ${B'}^{a_1\dots a_5}$, which do not enter in $A$, are
also absent in $H'_9$, so that we can ignore them;
(c) the field equations of $\omega^{ab} $ can be used to eliminate the $\omega^{ab}$;
(d) the linearized field equations for the elfbein $e^a$ and the gauge
one-form fields $B^{ab}$ and ${B}^{a_1\dots a_5}$ have a
structure similar to the  $e^a$ equation of $D=11$ supergravity and
(e) the linearized field equations for $\psi^\alpha$ and ${\psi'}^\alpha$
are linearized Rarita-Schwinger equations. With these assumptions, the counting of
on-shell degrees of freedom goes as follows:
\begin{equation}
\label{counting}
   \left(
     \begin{array}{cccc}
       e^a_\mu & (\psi^\alpha_\mu\, , \; {\psi'}^\alpha_\mu) & B^{ab}_\mu & B^{a_1\dots a_5}_\mu \\
       9\cdot 11-55 & \frac{32\cdot 8}{2}\, \textrm{each} & 9\cdot {{11}\choose{2}} & 9 \cdot {{11}\choose{5}} \\
     \end{array}
   \right)\ ,
\end{equation}
{\it i.e.} there are 4697 bosonic  and 256 fermionic degrees of
freedom\footnote{The vielbein $e^a_\mu$ and
Rarita-Schwinger $\psi^\alpha_\mu$ fields in $D$ dimensions
have, respectively, $(D-2)D-{D \choose {2}}= \frac{1}{2}(D-1)(D-2)$$-$1
(after using local Lorentz invariance) and $\frac{1}{2} (D-3) 2^{[D/2]}$
on-shell degrees of freedom. Similarly, a $p$-{\it form} gauge field
$A_{\mu_1\dots\mu_p}$ has ${D-2 \choose p}$ on-shell $d.o.f.$; the $B$'s
above are one-form gauge fields with additional antisymmetric $a$ indices.}.
But $D=11$ supergravity has 44+84=128  bosonic and 128
fermionic degrees of freedom, so that for $H'_9$ to lead to CJS supergravity
there should be 128 fermionic and 4569 bosonic extra hidden gauge symmetries.

Secondly, there is no reason why the $H'_9$ term in the expansion
\eqref{lambdaexp11} of the right dimension $L^9$ should correspond to a
locally supersymmetric action. Besides, the local supersymmetry transformations
of $D=11$ supergravity are not $osp(1|32)$ gauge transformations, but rather
local superspace transformations of the component fields the commutators of
which close on-shell only (see, for instance, \cite{DFR91}). It is thus unclear
how the $osp(1|32)$ gauge transformations could lead to these local superspace
transformations after selecting the  $H'_9$ term in \eqref{lambdaexp11}.

A second problem is the three-form field $A$ in the action of
CJS supergravity. For $A$ to be a composite field,
$A=A(e^a, B^{ab}, B^{a_1\dots a_5},\psi^\alpha, {\psi'}^\alpha)$,
the supersymmetry algebra of the $H'_9$ term in \eqref{lambdaexp11} would have to be
related with the algebra defined by the MC equations including the
one-form gauge fields appearing in the expression of a composite $A$.
A natural candidate for a supersymmetry algebra would be a contraction
of $osp(1|32)\oplus \widetilde{osp}(1|32)$ but, as shown in \cite{FIO15},
there is no way of obtaining by contraction the algebras given
in \cite{DF82, BAIPV04} that allow for a one-forms decomposition
of the CJS supergravity three-form field $A$.

As we have seen, already in the $D=5$ case where there is no $A$
complicating matters, the CS action does not lead to $D=5$ supergravity. So it is hard
to imagine why moving to $D=11$ would improve the situation so that supersymmetry is
preserved after selecting the proper $H'_9$ term in the expansion  \eqref{lambdaexp11}.
Further, if there were such a mechanism, working only in $D=11$ and ensuring
local supersymmetry after taking a non-leading term, it would presumably also
apply to the $H'_{10}$ and $H'_{11}$ terms in \eqref{lambdaexp11}; again,
this  would yield a $D=11$ supergravity with a cosmological
constant, which has been shown not to exist \cite{BDHS97}.

  The $D=11$ case is more convoluted than the $D$=5 one not only due to the
three-form field $A$, but also because of the auxiliary zero-form field
$F_{a_1\cdots a_4}$ which has to be added in the
first order formulation of $D=11$ supergravity, which is the one that
would naturally appear from a CS action. But even if these
difficulties were overcome, the $D$=5 case already tells us that the
resulting action would not be locally supersymmetric. In fact, an attempt made
in \cite{IR12} using just one $osp(1|32)$ algebra, ignoring $A$ and $F_{a_1\cdots a_4}$
and keeping only $e_a$, $\omega_{ab}$ and $\psi^\alpha$, supports this conclusion.

One may consider adding separately an $osp(1|32)$-gauge invariant
dimensionless three-form field $\mathcal{A}$ to look for an action
involving the fields of a single $osp(1|32)$. The additional
$\mathcal{A}$ is inert under $osp(1|32)$ gauge transformations and,
under two-form gauge transformations  $\Lambda$, $\mathcal{A}$ transforms as
$\delta_\Lambda \mathcal{A} = d \Lambda$; thus, the four-form $d\mathcal{A}$ is
$\delta_\Lambda$-gauge invariant. Then, the general gauge invariant
twelve-form $H(\mathbb{F}, \mathcal{A})$ ({\it cf.} \eqref{H11}) is given by
\begin{eqnarray}
\label{f3}
\nonumber H & = & Tr(\mathbb{F}^6) + \alpha Tr(\mathbb{F}^4) Tr(\mathbb{F}^2)+ \beta
\left( Tr(\mathbb{F}^2) \right)^3 + \nu Tr(\mathbb{F}^4)d\mathcal{A} \\
& + &\delta\left( Tr(\mathbb{F}^2) \right)^2d\mathcal{A} +\rho Tr(\mathbb{F}^2)(d\mathcal{A})^2 +
\sigma (d\mathcal{A})^3\ ,
\end{eqnarray}
where $\alpha,...,\sigma$ are dimensionless constants.

An action with the right dimensions would correspond to the $H_9$ term in the expression above
with $\mathcal{A}=\lambda^3 A$, $[A]=L^3$. However, this construction still would not explain
the need for the auxiliary $F_{a_1\cdots a_4}$ fields. In fact, one of the results of this
paper is that, since contractions do not appear to play a role in the present
problem, the field re-scalings need not being those that allow for
a consistent $\lambda\rightarrow 0$ limit.
Once $f_a=\lambda e_e$ is chosen, consistency of the contraction
limit would  require a new field, $e_{a_1\cdots a_5}$ say,  with $f_{a_1\cdots a_5}=
\lambda B_{a_1\cdots a_5}$, so that the $osp(1|32)$ MC equations
\begin{equation}
\label{MCcontr}
    df^a \propto \epsilon^{ab_1\cdots b_5 c_1\cdots c_5} f_{b_1\cdots b_5} f_{c_1\cdots c_5}+ \cdots
\end{equation}
have a well defined $\lambda \rightarrow 0$ limit. But, if this consistency condition is removed,
we may now set $f_{a_1\cdots a_5}= \omega_{a_1\cdots a_5}$, $[\omega_{a_1\cdots a_5}]= L^0$,
(rather than $f_{a_1\cdots a_5}= \lambda B_{a_1\cdots a_5}$, which implies $[B_{a_1\cdots a_5}]=L^1$).
Indeed, it will be shown that the $\omega_{a_1\cdots a_5}$ fields play
the role of the $F_{a_1\cdots a_4}$ (see below eq.~\eqref{eq7}).
Unfortunately, a calculation shows that the $\lambda^9$ term in the expansion of this new,
generalized CS action is not $D=11$ supergravity (in particular, the fermion
equation will not correspond to the spinor equation for CJS supergravity).
This was to be expected since, again, there is no reason for this term to
be invariant under supersymmetry gauge transformations.

   Nevertheless, we will show below that our construction for the fields
associated with the bosonic part of a  $osp(1|32)$,
supplemented by the three-form  $\mathcal{A}$,
does work for the {\it bosonic} sector of $D=11$ supergravity. In other
words, there are constants $\alpha,\cdots , \sigma$ in \eqref{f3} such
that the $H_9$ term in $H$ resulting from the re-scalings $f^a=\lambda e^a$,
$f^{ab}= \omega^{ab}$, $f^{a_1\cdots a_5}= \omega^{a_1\dots a_5}$ and
$\mathcal{A}=\lambda^3 A$ lead to the equations of its  bosonic sector.
In particular, the $\omega^{a_1\dots a_5}$ equation determines $\omega^{a_1\dots a_5}$
itself in terms of the coordinates of $dA=(dA)_{a_1\cdots a_4}e^{a_1}\cdots e^{a_4}$,
\begin{equation}
\label{omegadA}
   \omega_{a_1\dots a_5} \propto (dA)_{[a_1\cdots a_4} e_{a_5]} \ ,
\end{equation}
so that $\omega_{a_1\dots a_5}$ plays the role of the auxiliary $zero$-forms
of $D=11$ supergravity. In this way, the fact that the $D=11$ supergravity
action contains a generalized `CS term' for the field $A$, the
eleven-form $A\,dA\, dA$,
is incorporated into the full bosonic action through the sum of powers of
$\lambda$ described above. This result also extends others in
refs.~\cite{IMPRS09,IR12} in which standard pure gravity with just
$\omega^{ab}$ and $e^a$, without the fields $\phi$ in $D=5$ and $A$
in $D=11$, is derived from a CS action in these odd dimensions.

The plan of the paper is as follows. The `generalized CS action' is
defined in Sec.~2, where its expression in powers of the scale
factor $\lambda$ is given. Then, we study in Sec.~3 the field equations of
the model and compare them with those of the bosonic sector of supergravity.
We end with some conclusions and further comments. Some
calculations are relegated to an Appendix.

\section{The generalized $sp(32)$ Chern-Simons action}
\subsection{ $sp(32)$ Cartan structure equations and  gauge transformations}
In terms of its MC forms
$f^{\alpha}_{\hspace{1,5mm}\beta}$, $\alpha,\beta=1,\dots 32$,
the $sp(32)$ algebra is defined by
\be
\label{MCeq}
df^{\alpha}_{\hspace{1,5mm}\beta} =
-f^{\alpha}_{\hspace{1,5mm}\gamma}\wedge f^{\gamma}_{\hspace{1,5mm}\beta}
\hspace{0,3cm}, \hspace{0,7cm} df = -f^2\hspace{0,3cm}.
\ee
Using the symplectic metric $C_{\alpha \gamma}= -C_{\gamma\alpha}$, $f_{\alpha\beta}$ is given by
\be
f_{\alpha\beta} = C_{\alpha \gamma} f^{\gamma}_{\hspace{1,5mm}\beta} \quad, \quad
f_{\alpha\beta} = f_{\beta\alpha} \quad.
\ee
Since $f_{\alpha\beta}$ is a $32\times 32$ symmetric matrix, it  can be
expanded in the basis of ($\alpha\beta$)-symmetric matrices given by
`weight one' antisymmetrized products of $D$=11 Dirac matrices as
\be
\label{f11}
f_{\alpha \beta} = f_a\gamma^a_{\alpha \beta} + \frac{1}{4} f_{ab}\gamma^{ab}_{\alpha \beta}
+f_{a_1 ...a_5}\gamma^{a_1 ...a_5}_{\alpha \beta}  \hspace{0,3cm}.
\ee
The $1/4$ factor is introduced to obtain the usual relation between the spin
connection and its curvature (eq.~\eqref{RLRab}) as well as the definition
of the torsion (eq.~\eqref{TL}).

Gauge curvatures are introduced by moving from the MC equations
(zero curvature) to the Cartan structure ones, in which the
$sp(32)$ curvatures express the failure of $f$ to satisfy
the $sp(32)$ algebra MC equations. Let $\Omega$ be the two-form
matrix incorporating the curvatures. Then,
\be
\Omega =\mathcal{D}f = df +f^2\hspace{0,2cm},
\ee
where $f$ contains the one-form gauge fields, and
\be
d \Omega = \Omega f -f \Omega = [\Omega, f]  \quad ,
\ee
is the Bianchi identity $\mathcal{D}\Omega= d\Omega+ [f, \Omega] \equiv 0$
for the $sp(32)$ connection $f$. As $f$, the curvature $\Omega$ may be
similarly expressed as
\be
\label{Omegagamma}
\Omega_{\alpha\beta} = \Omega_a \gamma^{a}_{\alpha\beta}
+ \frac{1}{4}\Omega_{ab} \gamma^{ab}_{\alpha\beta} +
\Omega_{a_1...a_5}\gamma^{a_1...a_5}_{\alpha\beta}
 \hspace{0,2cm}.
\ee
The infinitesimal gauge transformations of $f,\Omega$ are given by
the standard expressions,
\be
\label{gautran}
\delta_b f = db + fb -bf = db + [f,b] \;\: , \;\;\delta_b \Omega = \Omega b - b\Omega = [\Omega,b] \ ,
\ee
where the zero-form matrix $b = b^{\alpha}_{\hspace{2mm}\beta}$ contains
the gauge functions
\be
b =  b_a \gamma^a + \frac{1}{4} b_{ab} \gamma ^{ab}
+ b_{a_1...a_5} \gamma ^{a_1...a_5} \hspace{0,2cm} .
\ee

\subsection{Generic expression for a CS-type action}
\label{genCS}
Since the bosonic sector of  $D = 11$ supergravity contains the three-form field $A$,
we add it explicitly to the one-form $sp(32)$ fields by introducing the
three-form $\mathcal{A}$ inert under $sp(32)$ $\delta_b$ gauge transformations
and under $\delta_\Lambda$ ones. Thus, the most general twelve-form $H(\Omega,\mathcal{A})$,
closed and invariant under both $\delta_b$ and $\delta_\Lambda$ gauge transformations,
may be written as
\begin{eqnarray}
\label{f3bis}
\nonumber H & = & Tr(\Omega^6) + \alpha Tr(\Omega^4) Tr(\Omega^2)+ \beta  \left( Tr(\Omega^2) \right)^3 +
\nu Tr(\Omega^4)d\mathcal{A} \\
& + &\delta\left( Tr(\Omega^2) \right)^2 d \mathcal{A}+\rho Tr(\Omega^2)(d \mathcal{A})^2 +
\sigma (d \mathcal{A})^3  \ ,
\end{eqnarray}
where the bosonic $\Omega$ has replaced $\mathbb{F}$ in eq.~\eqref{f3}, in which
fermions were present. Then, the integral
\be
\label{IGCS}
I =\int_{\mathcal{M}^{11}}    B, \quad dB=H\ ,
\ee
may be used to obtain a CS-type action.

Our task now is to extract from eq.~\eqref{f3bis} the physically relevant terms (it will turn
out that only the first term $Tr(\Omega^6)$ and those in $\nu$ and $\sigma$ will contribute)
and to fix their corresponding coefficients so that the resulting action determines the equations
of motion for the bosonic sector of supergravity. Because of the presence of
the three-form  $\mathcal{A}$, this action will be referred to as the {\it generalized}
CS action for the bosonic sector of $D$=11 supergravity.

\subsection{Generalized CS action for the bosonic sector of $D$=11  supergravity}
Again, the component fields in the one-form $f$, the two-form
$\Omega$ and the three-form $\mathcal{A}$ field
are dimensionless. Dimensions are introduced by setting
\be
\label{iden2}
\mathcal{A} = \lambda^3 A \hspace{0,2cm},\hspace{0,3cm}  \left[ A \right] = L^3 \hspace{0,3cm} ,
\ee
\be
\label{iden1}
f =  \lambda\, e_a\gamma^a + \frac{1}{4} \omega_{ab} \gamma^{ab}
+ \omega_{a_1...a_5}\gamma^{a_1...a_5}\hspace{0,3cm},
\ee
where in \eqref{f11} we set
\begin{eqnarray}
\nonumber f_a  &=&  \lambda e_a \hspace{0,3cm},		 \hspace{1,5cm} \left[e_a\right] = L \hspace{0,3cm}, \\
 f_{ab} &=& \omega_{ab}  \hspace{0,3cm},			\hspace{1,5cm} \left[\omega_{ab}\right] =
 L^0 \hspace{0,3cm},\\
\nonumber f_{a_1...a_5} &=& \omega_{a_1...a_5}  \ , \hspace{1,1cm}  \left[\omega_{a_1...a_5}\right] = L^0 \ .
\end{eqnarray}
With our mostly plus metric we use real gamma matrices such that
$\gamma^{a_1\dots a_{11}} = \epsilon^{a_1\dots a_{11}}$. Besides
the $1/4$ factor in \eqref{iden1} that
was fixed in \eqref{f11}, there is no special reason for the factors
accompanying the fields $e_a$, $\omega_{a_1\cdots a_5}$ and $A$.
Different coefficients would lead to different values for
the constants $\alpha,\dots ,\sigma$ in \eqref{f3bis} after requiring that the
action corresponds to the bosonic sector of supergravity. Thus, these
constants depend on the way the fields are introduced and will not affect
the final result. Keeping this in mind, we now look for the relevant terms
and their coefficients for the particular choices in \eqref{iden1}, \eqref{iden2}.

An action for $D$=11 gravity has dimensions $L^{D-2}$=$L^9$.
Thus, writing now $H|_i$ for $H_i$ and expressing the twelve-form
$H$ in \eqref{f3bis} and the eleven-form $B$ in powers of $\lambda$,
we obtain
\begin{eqnarray}
\nonumber
H &=&  H |_0 + \lambda \, H |_1 +... \, , \\
B &=&  B |_0 + \lambda \, B |_1 +... \ .
\end{eqnarray}
Then, $H|_i=dB|_i$ allows us to write for the different
$I_{GCS}|_i=\int_{\mathcal{M}^{11}} B|_i$,
\be
I_{GCS} = I_{GCS} |_0 + \lambda \, I_{GCS} |_1 +...  \  .
\ee
The physically relevant term is in  $\lambda^9$ since $\left[  I_{GCS} |_9  \right] = L^{9}$.
Therefore, $I_{GCS} |_9 = \int_{\mathcal{M}^{11}} B |_9$.

We are thus interested in $H|_9$. Since $H$ contains the $sp(32)$
curvature two-forms $\Omega_a$, $\Omega_{ab}$, $\Omega_{a_1...a_5}$
of \eqref{Omegagamma}, we need their expressions in terms of $e_a$, $\omega_{ab}$,
$\omega_{a_1...a_5}$.
To simplify the calculations, we write
\be
\label{Odff2}
\Omega = df + f^2 = \Omega_0 + \lambda \Omega_1 + \lambda^2 \Omega_2 \;  ,
\ee
with $f$ in \eqref{iden1} expressed as
\be
\label{Ome}
f = \lambda e + \omega_L +\omega_5 = \lambda e + \omega   \  ,
\ee
where $e=e_a \gamma^a$, $\omega_L=\frac{1}{4}\omega_{ab}\gamma^{ab}$ is the spin connection,
$\omega_5=  \omega_{a_1...a_5}\gamma^{a_1...a_5}$ and  $\omega = \omega_L + \omega_5 $.
In this way, the $sp(32)$-valued curvature in \eqref{Odff2} gives
\begin{eqnarray}
\label{Omegal}
\Omega &=& d(\lambda e   + \omega) + (\lambda e + \omega)(\lambda e + \omega)   \nonumber \\
&=& d\omega + \omega^2 + \lambda(de + \omega e+ e\omega ) + \lambda^2e^2 \nonumber \\
&\equiv&  R(\omega) + \lambda T + \lambda^2\Omega_2  \  .
\end{eqnarray}
Thus, $\Omega_0=R(\omega)=d\omega+ \frac{1}{2}[\omega,\omega]$,
$\Omega_1= T(e,\omega)= de +[\omega,e]$ and $\Omega_2(e)=e^2=\frac{1}{2}[e,e]$.
Notice that $T$ contains a piece proportional to $\gamma^a$ and another
proportional to $\gamma^{a_1...a_5}$; similarly, the curvature $R(\omega)$
contains contributions proportional to $\gamma^a$, $\gamma^{ab}$ and
$\gamma^{a_1...a_5}$, because it depends on both $\omega_L$ and $\omega_5$.
The previous equations tell us that to obtain the piece $H|_9$ that comes {\it e.g.}
from $Tr(\Omega^6)$, one has to  consider all the contributions containing  a number
$n_0$ of $R$ factors, $n_1$ of $T$ and $n_2$ of $\Omega_2$ in such a way that
\ben
\item$n_0 + n_1 + n_2 = 6$  (there are 6 curvatures)
\item$n_1 + 2n_2 = 9$ \ ,
\een
where the first condition guarantees that the order of the forms is
twelve and the second one that their length dimension is nine.
The only two solutions are:
\bi
\item $n_2 = 4$, $n_1 = 1$, $n_0 = 1$ , or \hspace{0,5cm}
\item $n_2 = 3$, $n_1 = 3$, $n_0 = 0$
\ei
Thus, the $R, T, \Omega_2$ contributions are of the
form
\begin{eqnarray}
\label{Hexpanded}
Tr(\Omega^6)|_9 = Tr(\mathcal{W}(\Omega^4_2, T, R))
+ Tr(\mathcal{W}(\Omega^3_2, T^3, R^0)) \hspace{0,2cm},
\end{eqnarray}
where {\it e.g.} $\mathcal{W}(\Omega^4_2, T, R)$ is the sum of all nine-form
`words' that can obtained out of four $\Omega_2$, one $T$ and one $R$.
This would give us the piece $Tr(\Omega^6)|_9$ of $H|_9$. We could now
add to \eqref{Hexpanded} the contributions to $H|_9$ coming from the
other terms in \eqref{f3bis}, to find an $11$-form $B|_9$ with $dB|_9=H|_9$,
and compare with the action of the bosonic sector of $D=11$ supergravity.
Instead, we will obtain directly the field equations for the action $\int_\mathcal{M} B_9$
from the original, unexpanded $H$ twelve-form.

\section{Field equations}
The field equations for $I_{GCS}$ can be obtained directly from $H$ in a way
similar to that used in \cite{AIMV}. To find them, we use the following fact (see \cite{maxw}):
let $i_{f^\alpha{}_\beta}$, $i_{\Omega^\alpha{}_\beta}$, $i_{\mathcal{A}}$ and
$i_{d\mathcal{A}}$ be the inner
derivations associated with the fields and curvatures of
the algebra with respect to $f$, $\Omega$, $\mathcal{A}$ and $d\mathcal{A}$, defined by
\begin{equation}
\label{inner}
     i_{f_{\alpha \beta}} f_{\gamma \delta} = \delta^\alpha_{(\gamma} \delta^\beta_{\delta)}\ ,
\quad i_{\Omega_{\alpha \beta}}  \Omega_{\gamma \delta}
= \delta^\alpha_{(\gamma} \delta^\beta_{\delta)}  \ , \quad i_{\mathcal{A}} \mathcal{A}
= 1 \ , \quad i_{d\mathcal{A}}
d\mathcal{A} = 1 \ ,
\end{equation}
and zero otherwise. If $H=dB$ is a form defined on this algebra that defines the action
through $I=\int B$, then the field equations for $I$ are given by
$i_{\Omega^\alpha{}_\beta} H= 0$  and $i_{d\mathcal{A}} H = 0$.
Let us denote the equations of motion for $f$ and $\mathcal{A}$ by
$E(f) = 0$ and $E(\mathcal{A}) = 0$ respectively. Then, using
\eqref{inner} in  \eqref{f3bis} we obtain
\begin{eqnarray}
\label{eq0}
 \nonumber E(f) &=& 6\Omega^5 + 4\alpha Tr(\Omega^2)\Omega^3 + 2\alpha Tr(\Omega^4)\Omega +
 6\beta Tr(\Omega^2)^2\Omega \\
 &+& 4\nu d \mathcal{A} \Omega^3  + 4\delta d \mathcal{A} Tr(\Omega^2)\Omega + 2\rho (d \mathcal{A})^2
 \Omega = 0 \hspace{0,2cm},
\end{eqnarray}
where $E(f)$ is a ten-form, and by
\be
\label{eq9}
E(\mathcal{A}) = \nu\, Tr(\Omega^4) + \delta\, (Tr(\Omega^2))^2
+ 2\rho \, (d \mathcal{A}) Tr(\Omega^2) +3\sigma
(d \mathcal{A})^2 = 0\hspace{0,2cm} ,
\ee
where $E(\mathcal{A})$ is an eight-form.

We have to extract now from the above the equations for $e$, $\omega$ ($\omega_L$
and $\omega_5$) and $A$ for the action $I_{GCS}|_9$. Proceeding as in
\cite{AIPV07}, where the equations for the dimensionful fields were
derived from those for the dimensionless ones by selecting the appropriate powers
of $\lambda$, they are given by
\begin{eqnarray}
\label{completeset}
 \nonumber E(e) &=&  (E(f)|_{9-1=8})|_{\gamma^{[1]}} \hspace{0,2cm}, \\
E(\omega) &=&  E(f)|_{9} \hspace{0,2cm},\ E(\omega_L)=\left(E(f)|_{9}\right)_{\gamma^{[2]}} \hspace{0,2cm},
\ E(\omega_5)=\left(E(f)|_{9}\right)_{\gamma^{[5]}} \\
 \nonumber E(A) &=&  E(\mathcal{A})|_{9-3=6} \hspace{0,2cm}
\end{eqnarray}
since $[e]=L^1$, $[\mathcal{A}]=L^3$, $[\omega]=L^0$, and
where the subscripts $\gamma^{[2,5]}$ refer to the contributions proportional to the
antisymmetrization of two and five $D$=11 gamma matrices respectively. Eqs.~\eqref{completeset}
constitute the complete set of equations of our bosonic model.

We have to find now $E(f)|_{8}$, $E(\omega_L)|_{9}$, $E(\omega_5)|_{9}$ and $E(A)|_{6}$
by taking into account that
\be
\nonumber \Omega R+ \lambda T + \lambda^2 \Omega_2  \;\; , \;\;
d\mathcal{A} =  \lambda^3 dA \ .
\ee

\subsection{Field equation for $\omega$}
We need to know the contributions of all terms in equation~\eqref{eq0}, namely
all the contributions containing  $n_2$ factors $\Omega_2$ , $n_0$ factors $R$
and $n_1$ factors $T$ in such a way that the order of the form is $10$ and its
dimension $L^9$. Then, we find that the
$\omega$ equation is given by the ten-form expression
\be
\label{bigeq}
E(\omega)=E(f)|_9=
6\, \mathcal{W}(\Omega^4_2, T) + 4\nu dA e^6 = 0  \; ,
\ee
where the first term comes from the first one in eq.~\eqref{eq0} and the other
comes from the $\nu$ term. Since $\omega=\omega_L+\omega_5$,
eq.~\eqref{bigeq} contains two different contributions,
one proportional to $\gamma^{a_1 a_2}$ from the first term that gives the equation for
$\omega_L$,  and another proportional to $\gamma^{a_1... a_5}$ that comes
from both terms and gives the equation for $\omega_{5}$. We consider them now.

The first ten-form in \eqref{bigeq} is
\be
\mathcal{W}(\Omega^4_2, T) = e^8 T + e^6Te^2 + e^4 T e^4 + e^2T e^6 + T e^8\hspace{0,3cm},
\ee
where $T$ is given (see \eqref{Omegal},\eqref{Ome}) by
\begin{eqnarray}
\label{TL}
T = de + [\omega, e]
= T_L + [\omega_5, e]  \ ; \  T_L= de + [\omega_L, e] \; ,
\end{eqnarray}
and the explicit expression for the torsion $T_L$ is
\begin{equation}
\label{expTL}
    T_L = T^a \gamma_a = (de^a + \omega^a{}_b e^b) \gamma_a \ .
\end{equation}
Then, the first term on the $l.h.s.$ of \eqref{bigeq} can be written as
\begin{equation}
\label{eq3}
\mathcal{W}(\Omega^4_2, T) =  \mathcal{W}((e^2)^2, T_L) +  \mathcal{W}(e^9, \omega_5)
\hspace{0,3cm}.
\end{equation}

\subsubsection{Equation for $\omega_L (\omega_{a b}$)}
To see how the $\gamma_{a_1 a_2}$ and $\gamma_{a_1 ... a_5}$ contributions come out,
note the identity
\be
\label{gsum}
\gamma_{a}\gamma_{a_1 ... a_k} = \sum^k_{i = 1} (-1)^{i-1} \,\eta_{a a_i}\gamma_{a_1 ... \hat{a}_i ... a_k} +
\gamma_{a a_1...a_k}\hspace{0,3cm}.
\ee
When contracted with the indices of, say $e^a B^{a_1...a_k}$, one gets:
\be
\label{afcon}
e^a \gamma_a B^{a_1...a_k}\gamma_{a_1...a_k} = k\, e^a B_{a a_2...a_k}  +
\gamma_{a a_1...a_k}e^a B^{a_1...a_k}  \hspace{0,2cm},
\ee
{\it i.e.}, all terms in the sum \eqref{gsum} add up, and the first term appears $k$ times.
The same pattern exists when two matrices $\gamma_{a_1...a_k}$, $\gamma_{a_1.....a_s}$
are multiplied, but now there are contributions with all possible number of contractions.
The $e^8T_L$ terms have the structure $\gamma^{[8]}\cdot\gamma^{[1]}$ (again, the superscripts
indicate the number of $\gamma$'s in the skewsymmetric products). This gives, schematically,
\be
\gamma^{[8]}\cdot\gamma^{[1]} \sim \gamma^{[9]}+ \gamma^{[7]}
\ee
where there are no contractions in $ \gamma^{[9]}$ and one in $ \gamma^{[7]}$.
The $ \gamma^{[7]}$ contribution will cancel because only the matrices
symmetric in all indices contribute ($\gamma^{[1,2,5,6,9,10]}$ are
symmetric; {\bf 1}, $\gamma^{[3,4,7,8]}$ skewsymmetric).
Thus, only the $e^8 T_L$ terms appear in the $\omega_L$ equation since
\be
\gamma^{a_1...a_9} \propto \epsilon^{a_1...a_9 a b} \gamma_{a b} \hspace{0,3cm}.
\ee

In general, since with our metric signature we can choose
$\gamma^{a_1...a_{11}} = \epsilon^{a_1...a_{11}}$, we have
\begin{equation}\label{geg}
\gamma^{a_{k+1}...a_{11}}= \frac{(-1)^{\frac{k(k-1)}{2}}}{k!} \epsilon^{b_1\dots b_k a_{k+1}...a_{11} }
\gamma_{b_1\dots b_k}\ .
\end{equation}
On the other hand, the terms $e^9 \omega_5$  coming from \eqref{eq3} are,
again schematically, of the form
\be
\gamma^{[9]}\cdot\gamma^{[5]} \sim \gamma^{[14]}+ \gamma^{[12]}+ \gamma^{[10]}
+ \gamma^{[8]}+ \gamma^{[6]}+ \gamma^{[4]}\hspace{0,3cm},
\ee
The $\gamma^{[10]}\sim \gamma^{[1]}$ contribution vanishes because there is no
$\omega_a$, {\it i.e.} there is no equation of dimension $L^9$ with a
single Lorentz index.  The only symmetric $\gamma$ is $\gamma^{[6]}$.
So the $e^9 \omega_5$ terms only appear in the $\omega_5$ equation.
The $\omega_L$ equations are then
\begin{equation}
\label{eq10}
(E(f)|_9)_{\gamma^{[2]}}= E(\omega_L) \propto
e_{a_1}...e_{a_8} (T_{L})_{a_9} \gamma^{a_1...a_9}= 0  \; .
\end{equation}
This equation implies $T_L = 0$, which, as usual, can be used to express
$\omega_{ab\mu}$ in terms of $e^a_{\hspace{0,2cm} \mu}$ and its derivatives.

\subsubsection{Equation for $\omega_5$ ($\omega_{a_1...a_5}$)}
This equation has contributions from the two terms in \eqref{bigeq}. One is given
by its second term $4\nu (dA) e^6$ which, due to $e^6$, is proportional to $\gamma^{[6]}$,
and the other is the contribution with four contractions from the terms
with nine $e$ and one $\omega_5$ from $\mathcal{W}(e^9, \omega_5)$,
which is also proportional to $\gamma^{[6]}\sim \gamma^{[5]}$, contained in
the first one, $6\,\mathcal{W}(\Omega^4_2, T)$.
A long calculation shows that this second contribution is given by
\be
  2\cdot \frac{9!}{4!} \, e_{a_1}... e_{a_5} e^{b_1}... e^{b_4}\omega_{b_4.....b_1a_6}
 \gamma^{a_1...a_6} \hspace{2mm}.
\ee
Taking into account both terms, the $\omega_5$ equation of motion is found to be
\begin{eqnarray}
\label{omega5}
&&(E(f)|_9)_{\gamma^{[5]}}=E(\omega_5) =\\
\nonumber
&&12\cdot \frac{9!}{4!} \, e_{a_1}... e_{a_5} e^{b_1}... e^{b_4}\omega_{b_4...b_1a_6}\gamma^{a_1...a_6}+
4\nu\, dA\,  e_{a_1}... e_{a_6} \gamma^{a_1...a_6} = 0 \, .
\end{eqnarray}

Let us see what this equation leads to. In terms of the elfbein components
of $dA$,
\be
\label{eq8}
dA = (dA)_{b_1...b_4}e^{b_1}...e^{b_4} \hspace{2mm} \; ,
\ee
it reads
\be
\label{otran}
\frac{9!}{2} \, e_{a_1}... e_{a_5} e_{b_1}... e_{b_4}e_c \omega^{b_4...b_1}{}_{a_6}{}^c
\gamma^{a_1...a_6} + 4\nu (dA)^{b_1...b_4} e_{a_1}... e_{a_6}e_{b_1}... e_{b_4}
\gamma^{a_1...a_6} = 0 \hspace{2mm},
\ee
where $\omega^{b_4...b_1}{}_{a_6}  = \omega^{b_4...b_1}{}_{a_6}{}^c e_c$.
We now write the products of ten $e$'s above  as
$$ e_{a_1}...e_{a_5}e_{b_1}...e_{b_4}e_{c}  =  \epsilon_{a_1...a_5b_1...b_4cd}E^d \hspace{2mm},$$
for some ten-form $E^d$. Then, factoring out this form in eq.~\eqref{otran}
and $\gamma^{a_1...a_6}$, we find
\be
\label{ec32}
\frac{9!}{2} \epsilon_{b_1...b_4cd\left[a_1...a_5\right.} \omega^{b_4...b_1}{}_{\left. a_6\right]}{}^{c} +
4\nu \epsilon_{a_1...a_6b_1...b_4d} (dA)^{b_1...b_4} = 0 \hspace{2mm},
\ee
where $\left[ \hspace{0,2cm} \right]$ indicates weight one antisymmetrization
in $a_1...a_6$. It is shown in the Appendix (sec.~\ref{app1}) that the solution is
\be
\label{c1}
\omega^{d_1...d_5}_{\hspace{8,5mm}d} = - \frac{40}{9!} \, \nu \, (dA)^{\left[d_1 ... d_4\right. }
\delta^{ d_5 \left. \right] }_d    \hspace{2mm}.
\ee
This equation relates the one-form gauge field components $\omega^{d_1...d_5}$ to
those of the four-form $F = dA$. It can also be written as
\be
\label{eq7}
\omega^{d_1...d_5} = - \frac{40}{9!} \, \nu \, (dA)^{[d_1 ... d_4 } e^{ d_5]}   \hspace{2mm}.
\ee
Hence, $\omega^{d_1...d_5}$ may be expressed in terms of the coordinates of $dA$ so
that, as anticipated, $\omega_5$  plays a role analogous to that of the auxiliary zero-forms
$F_{a_1\cdots a_4}$ of the first order formulation of $D=11$ supergravity,
where $F\propto dA$.
\bigskip

\subsection{Field equation for $A$}
The sum of the contributions to the field equation~\eqref{eq9} with the
right dimension, $E(\mathcal{A})|_6 =E(A)=0$ (see \eqref{eq9}), leads to
\be
\label{Aeq}
 E(A)= 4\, \nu \, 32 e_{a_1}...e_{a_6}\, D \, \omega_{{a_7}...a_{11}} \epsilon^{a_1 ... a_{11}}+
 3 \sigma (dA)^2 = 0 \hspace{2mm},
\ee
where again $D$ is the $\omega_L$ covariant derivative; we see that there is no
contribution from the $\delta$ and $\rho$ terms. In the $e^a$ basis, this gives
\be
\nonumber 4\,  \nu \, 32 e_{a_1}...e_{a_6}\, D_{b_1}  \omega_{{a_7}...a_{11}b_2}e^{b_1}e^{b_2}
\epsilon^{a_1 ... a_{11}} = -3\, \sigma (dA)_{b_1 ... b_4}(dA)_{c_1 ... c_4}
e^{b_1}...e^{b_4}e^{c_1}...e^{c_4}\hspace{2mm}.
\ee

Now we can introduce the eight-form
$E_{d_1 d_2 d_3} \equiv \epsilon_{d_1 d_2 d_3 b_1 ... b_8}e^{b_1}...e^{b_8}$,
and use it to rewrite the factors with eight one-forms $e^a$.
If the $E_{d_1 d_2 d_3}$ are then factorized, we obtain
\be
\label{eqAcoord}
 4\,  \nu \, 32 \, \cdot 6! \,\delta^{a_7 ... a_{11}}_{b_1 b_2 d_1 ... d_3}\,
D^{b_1} \,  \omega_{{a_7}...a_{11}}^{\hspace{1cm}b_2} = 3\, \sigma \epsilon_{b_1 ... b_4 c_1 ... c_4 d_1 ... d_3}
(dA)^{b_1 ... b_4}(dA)^{c_1 ... c_4} \hspace{2mm}.
\ee
Using the expression \eqref{c1} for  $\omega_{{a_7}...a_{11}}^{\hspace{1cm}b_2} $ in terms of the
components of $dA$, the $r.h.s$ of \eqref{eqAcoord} reads
\begin{eqnarray}
\nonumber
\delta^{a_7 ... a_{11}}_{b_1 b_2 d_1 d_2 d_3}\, D^{b_1} \,  \omega_{{a_7}...a_{11}}^{\hspace{1cm}b_2}
&=& -\frac{40}{9!}\, \nu \, \delta^{a_7 ... a_{11}}_{b_1 b_2 d_1 d_2 d_3}
\delta^{b_2}_{a_{11}} \, D^{b_1}\, (dA)_{a_7 ... a_{10}} \\
\nonumber
&=& -\frac{40}{9!}\,\nu \, \delta^{a_7 ... a_{10}b_2}_{b_1 b_2 d_1 d_2 d_3}  \, D^{b_1}\,
(dA)_{a_7 ... a_{10}} \\
\nonumber
&=& -(-7)\, \frac{40}{9!}\,\nu \, \delta^{a_7 ... a_{10}}_{b_1  d_1 d_2 d_3}  \, D^{b_1}\,
(dA)_{a_7 ... a_{10}} \\
\nonumber
&=& 4!\cdot7\,\frac{40}{9!}\,\nu  \, D^{b_1}\, (dA)_{b_1 d_1 d_2 d_3}  = \frac{1}{54} \,
\nu  \, D^{b_1}\, (dA)_{b_1 d_1 d_2 d_3} \hspace{2mm}.
\end{eqnarray}
In this way, the final expression for the $A$ equation of the motion is found to be
\begin{eqnarray}
\label{f1}
 D^{b_1}\, (dA)_{b_1 d_1 d_2 d_3} = \left(\frac{9 \,\sigma}{5120\, \nu^2}\right)
 \epsilon_{b_1 ... b_4 c_1 ... c_4 d_1 d_2 d_3}(dA)^{b_1 ... b_4}(dA)^{c_1 ..... c_4}
 \hspace{2mm}.
\end{eqnarray}
Note that this equation has the form required to reproduce the equations of
$D=11$ supergravity in the absence of fermions (see \cite{Silva,DFR91,CJS}).
\bigskip

\subsection{Field equation for $e$}
We need to know the contributions of all terms in eq.~\eqref{eq0} again,
but now we have to find ($E(f)|_8)|_{\gamma^{[1]}}=E(e)$ instead of $E(f)|_9$ in eq.~\eqref{completeset}.
Collecting all the possible contributions as explained before, we find
that they all come from the first and the $\nu$ term in eq.~\eqref{eq0},
\begin{eqnarray}
\label{eqea}
E(e) &=&
6  \mathcal{W}(\Omega^3_2, T^2)|_{\gamma^{[1]}} + 6 \mathcal{W}(\Omega^4_2, R)|_{\gamma^{[1]}}
\nonumber \\
 & & + 4 \nu dA (\Omega^2_2 T +
\Omega_2 T \Omega_2 + T \Omega_2^2)|_{\gamma^{[1]}} = 0\ ,
\end{eqnarray}
where, again,  $|_{\gamma^{[1]}}$ selects the contribution accompanying
a single gamma matrix $\gamma^a$, or equivalently, a ten indices gamma
matrix, $\gamma^{a_1 ... a_{10}}$.
In particular we need the contributions coming from the term
$6\mathcal{W}(\Omega^3_2, T^2)|_{\gamma^{[1]}} + 6 \mathcal{W}(\Omega^4_2, R)|_{\gamma^{[1]}}$
in \eqref{eqea}, but this  is a very tedious calculation.
Instead, it is more convenient to take advantage
of the fact that the symmetry of the stress-energy tensor forces its terms to
be the result of contracting three or four indices
among two $dA^{\mu \nu \rho \sigma}$ (in the $dx^{\mu}$ basis), namely
$(dA)^{\mu \nu \rho}_{\hspace{0,5cm}\alpha} \, (dA)_{\mu \nu \rho \beta}$ and
$(dA)^{\mu \nu \rho \sigma} \, (dA)_{\mu \nu \rho \sigma} \, g_{\alpha \beta}$.
Hence, Einstein's equations have the form
\be
\label{eq3.43}
R(\Gamma)_{\mu \nu} - \frac{1}{2}g_{\mu \nu}R(\Gamma) =
P \, (dA)^{\alpha \rho \gamma}_{\hspace{0,5cm}\mu}(dA)_{\alpha \rho \gamma \nu} +
Q\, (dA)^{\alpha \rho \gamma \delta} (dA)_{\alpha \rho \gamma \delta} g_{\mu \nu} \hspace{2mm},
\ee
with $P,Q$ yet to be determined. With the sign
for the curvature tensor as in \cite{FiPa:03}, $R(\Gamma)$ and $R(\omega_L)$ are related
through the elfbein postulate by $R(\Gamma) = 2 R(\omega_L) $.

The  $P$, $Q$ constants are now determined using that the covariant
derivative of the Einstein tensor is zero, $\nabla^{\mu}
\left( R(\Gamma)_{\mu \nu}-\frac{1}{2} g_{\mu \nu} R(\Gamma) \right)=0$.
Then, the $r.h.s$ of \eqref{eq3.43} must vanish when the supergravity field
equation for the $A$ field (equivalent to our (eq.~\eqref{f1})),
\be
\label{eq3.45}
\nabla^{\mu} (dA)_{\mu \nu \rho \sigma} \propto
\epsilon_{\nu \rho \sigma \lambda_1... \lambda_4 \tau_1...\tau_4}
(dA)^{ \lambda_1... \lambda_4} (dA)^{\tau_1...\tau_4} \ ,
\ee
where the proportionality factor is unimportant here, and
\be
\label{eq3.44}
\partial_{\left[\mu \right.}\, (dA)_{\left. \nu \rho \gamma \tau \right]}= 0\hspace{2mm}
\ee
($d(dA)\equiv 0$), are used.  Indeed, the covariant derivative of the
$r.h.s$ of eq.~\eqref{eq3.43} may be written using \eqref{eq3.44} as a linear combination of
$(dA)^{\rho\sigma\lambda\tau} \nabla_\nu (dA)_{\rho\sigma\lambda\tau}$ and
$(dA)_{\rho\sigma\lambda\nu} \nabla_\mu (dA)^{\rho\sigma\lambda\mu}$.
This last contribution vanishes due to eq.~\eqref{eq3.45}. Hence, the
first contribution also has to vanish and, since it includes
a factor $(P+8Q)$, it follows that $P/Q = - 8$ (see, {\it e.g.}, \cite{Ortin}).
Thus, we only need now the overall factor.

To fix it, we take the trace of eq. \eqref{eq3.43} to find the Ricci scalar
\begin{equation}
\label{eq3.46}
R(\Gamma)^{\mu\nu}{}_{\mu\nu}=
\frac{P}{12} \, (dA)_{\mu \nu \rho \sigma}\, (dA)^{\mu \nu \rho \sigma} \  .
\end{equation}
We still need the value of $P$ for our action.
If we compute the trace of the $E(e)=0$ (eq.~\eqref{eqea}) times $e^a\gamma_a$,
we obtain
\begin{eqnarray}
\label{riccieq}
0 &=& 6 \, Tr(9\omega_5 e \omega_5 e^8 + 9 \omega_5^2 e^9
+ 9\omega_5 e^2 \omega_5 e^7 + 9\omega_5 e^3 \omega_5 e^6 \\ \nonumber
&+& 9\omega_5e^4 \omega_5 e^5 )  + \frac{30}{4}Tr(  R_L e^9 )
+ 4\nu (dA) \,  6 Tr (\omega_5 e^6 )\hspace{2mm},
\end{eqnarray}
where the curvature $R_L$ is
\begin{equation}
\label{RLRab}
   R_L(\omega_L)= d\omega_L +\omega_L \, \omega_L
   = \frac{1}{4} (d\omega_{ab}+\omega_a{}^c\omega_{cb}) \gamma^{ab}
   \equiv \frac{1}{4} R(\omega_L)_{ab}\gamma^{ab}  \ .
\end{equation}
This expression leads to an equation for the Ricci scalar $R(\omega_L)^{ab}{}_{ab}$
that has the advantage that the different contributions are easier to compute. A calculation
(Appendix, eq.~\eqref{RoFF}) shows that $\nu$ in our action is related to $P$ by
\be
\label{valueP}
P = 12 \cdot 32 \cdot  \frac{4!\cdot 7!}{( 9!)^2} \nu^2 \ .
\ee
Now, to complete the $E(e)=0$ equation of supergravity we need to fix the
value of $\nu$ in \eqref{f3bis}, \eqref{eqea}; to determine
the equation $E(A)=0$ in \eqref{f1} we further
require the value of $\sigma$.

\subsection{The generalized CS action for the bosonic sector of $D = 11$ supergravity}
Having found the field equations from our action, we now fix the remaining
constants in \eqref{f3bis} so that the equations of bosonic $D=11$ supergravity
follow from $I_{GCS}$ as stated.
First, the $D$=11 supergravity equation for the $e$  field is,
after taking the trace (see e.g. \cite{FiPa:03}),
\be
\label{Rg}
R(\Gamma) = \left( \frac{1}{12} \right)^2\,  (dA)_{a_1\dots a_4} (dA)^{a_1\dots a_4} \hspace{2mm}.
\ee
Comparing with \eqref{eq3.46} we find $P=\frac{1}{12}$,
which in eq.~\eqref{valueP} then gives
\be
\label{nucuad}
\nu^2 =\left(\frac{1}{12}\right)^2 \frac{(9!)^2}{ 32\cdot 4! \cdot 7!}  \;
\ee

Secondly, the $D$=11 supergravity equation for $A$ is
\be
\label{easugra}
 D^{b_1}\, (dA)_{b_1 d_1 ... d_3} = \left(\frac{1}{3^2 \cdot 2^7}\right)
 \epsilon_{b_1 ... b_4 c_1 ... c_4 d_1 ... d_3} (dA)^{b_1 ... b_4}(dA)^{c_1 ..... c_4} \hspace{2mm}.
\ee
Comparing with our \eqref{f1} it follows that
\be
\label{relat}
\sigma = \nu^2 \left( \frac{40}{81}\right) \ .
\ee
The value of $\sigma$ follows using eq.~\eqref{nucuad} in eq.~\eqref{relat},
\be
\label{sigval}
\sigma =  \frac{5}{4\cdot (12)^2}\cdot  \frac{(8!)^2}{ 4!\cdot 7!} \hspace{2mm} .
\ee
Thus, the needed values of $\nu$ and $\sigma$ in \eqref{f3bis} are now
fixed; the terms in $\alpha, \beta,\delta, \rho$ do not appear once the relevant
$H_9$ term is selected. Note that it is possible to obtain $\nu$ from
\eqref{nucuad} because its r.h.s. is positive.

Summarizing, the  generalized CS action for the bosonic
sector of D=11 supergravity is obtained from
\be
\label{HforbosCS}
H  =  Tr(\Omega^6) + \nu Tr(\Omega^4)d\mathcal{A} +
\sigma (d \mathcal{A})^3  \; ,
\ee
with $\nu$ and $\sigma$ given by eqs.\eqref{relat} and \eqref{sigval}.
After the rescalings \eqref{iden2} and \eqref{iden1}, the action follows
from $B_{|9}$ with $dB_{|9}=H_{|9}$ and the equations of motion for the
$\omega$, $A$ and $e$ fields are given by eqs.~\eqref{bigeq}
[eqs.\eqref{eq10},\eqref{omega5}], \eqref{Aeq} and \eqref{eqea}
[\eqref{eq3.43}] respectively, the constants of which
have already been fixed. These equations are those of $D=11$ supergravity
when spinors are ignored, and hence $B_{|9}$ determines the generalized
CS action of its bosonic sector.

\section{Conclusions}
We have shown that the bosonic sector of $D=11$ supergravity may be obtained from a
generalized  CS action based on the one-form gauge fields of the
$sp(32)$ subalgebra of $osp(1|32)$ supplemented with a dimensionless
three-form field $\mathcal{A}$.
The need for $\mathcal{A}$ could not have been guessed without having in
mind $D=11$ supergravity: the presence of fermions requires $\mathcal{A}$
by simply counting the degrees of freedom of the $D = 11$ supermultiplet.
Further, we  have also shown (see \eqref{c1}) that the role of the
auxiliary zero-form fields $F_{a_1\dots a_4}$ that appear in the first-order
version of $D = 11$ supergravity \cite{Silva} is played by
specific gauge fields associated with $sp(32)$.

The values of the constants that determine our generalized
CS bosonic action were obtained by requiring that the equations
it leads to are those of the bosonic sector of $D = 11$ supergravity.
It turns out that only three terms in eq.~\eqref{f3bis} are actually needed,
the first one and those in $\nu$ and $\sigma$, since the others do not appear
in the bosonic equations obtained from the $\lambda^9$ term in the $\lambda$
expansion. The other terms and their constants would appear when including
fermions, eq.~\eqref{f3}, but nevertheless (Sec.~\ref{prelim}) this will not
lead to $D$=11 supergravity. Hence, there is no generalized CS action based on
$osp(1|32)$ with the addition of the three-form field leading
to CJS supergravity. Therefore, although $D=3$ supergravity may be
described by a CS action, we conclude that this is not so in larger,
odd spacetime dimensions.

It was already conjectured in the original paper \cite{CJS} that $osp(1|32)$
would provide the lead for a geometric interpretation of $D$=11 supergravity.
The main obstacle to relate its field contents to the geometric MC fields
of a superalgebra in the search for a possible CS action is the appearance
of the three-form field $A$.  As mentioned, it is possible to retain only
one-form fields by assuming a composite nature for $A$ \cite{DF82}
and then using a superalgebra that incorporates the one-form MC components
of $A$. In fact, there is a whole family of superalgebras related to $osp(1|32)$
that do just this \cite{BAIPV04} (another family of algebras structure has recently
been shown to exist for $N$=2, $D$=7 supergravity \cite{adar16}).

Summarizing, we have shown  that although there is no CS action for
CJS supergravity, its bosonic sector may be described by
a generalized CS action in the sense of
Sec.~\ref{genCS}. But, if we insist in including fermions,
we conclude that the only geometric way of relating CJS supergravity
to the $osp(1|32)$ superalgebra requires assuming the
mentioned composite nature for $A$ \cite{DF82,BAIPV04}.
Even so, the connection with $osp(1|32)$ is rather subtle \cite{BAIPV04}:
the family of algebras that trivialize the three-form $A$ are deformations of
an algebra which is the expansion $osp(1|32)(2,3)$  of $osp(1|32)$
in the sense of \cite{AIMV, AIPV07}.
\bigskip

\noindent
{\bf Acknowledgements}. This work has been partially supported by the
grants  MTM2014-57129-C2-1-P from the MINECO (Spain), VA057U16 from the Junta de
Castilla y Le\'on and by the Spanish Centro de Excelencia Severo Ochoa
Programme (IFIC-SEV-2014-0398). Correspondence with R. D'Auria is
also appreciated.

\noindent
\section{Appendix}
\label{av2}
This Appendix provides details of some main text calculations.
\subsection{Solving for $\omega_5$ in the $\omega_5$ equation}
\label{app1}
Let us solve \eqref{ec32} for $\omega_5$. Contracting the equation with
$\epsilon^{a_1...a_6d_1...d_5}$ we find,
\be
\frac{9!}{2} \epsilon^{ a_1...a_5 d_1...d_5} \, \epsilon_{ a_1...a_5 b_1...b_4 c d}
\omega^{b_4...b_1  \hspace{2mm}c}_{\hspace{8mm}a_6} + 4\nu \, \epsilon_{a_1.....a_6b_1...b_4d}\,
\epsilon^{ a_1...a_6 d_1...d_5} (dA)^{b_1...b_4} = 0 \hspace{2mm}.
\ee
Taking into account that
\be
\epsilon^{ a_1...a_k b_1...b_{11-k}}\epsilon_{ a_1...a_k c_1...c_{11-k}} = - k! \,
\delta^{b_1...b_{11-k}}_{c_1...c_{11-k}}
\ee
for our signature choice, $(- + ... +)$, where $\delta^{b_1...b_{k}}_{a_1...a_{k}} = \sum_{\sigma \in s_k}
\delta^{b_1}_{a_{\sigma(1)}}...\delta^{b_k}_{a_{\sigma(k)}} \hspace{2mm}$, we obtain
\be
\frac{9!}{2}5! \,  \delta^{a_6 d_1...d_{5}}_{b_1...b_{4}c d} \omega^{b_4...b_1
\hspace{2mm}c}_{\hspace{8mm}a_6} + 4\nu \, 6!\, \delta^{d_1...d_{5}}_{b_1...b_{4}d} (dA)^{b_1...b_4} = 0 \hspace{2mm}.
\ee
Now, using
\be
\delta_{b_1...b_{k}}^{a a_1...a_{k-1}} = \sum^k_{l = 1} \delta^{a}_{b_l}
\delta^{a_1 ... a_k}_{b_1 ... \hat{b}_l... b_k}
\ee
in the first term with $a = a_6$, it follows that
\be
\label{eq4}
\left( \frac{9!}{2}\, 5! \,  \omega^{b_4...b_1  \hspace{2mm}c}_{\hspace{8mm}c}
+ 4\nu \, 6!\,   (dA)^{b_1...b_4} \right)
\delta^{d_1...d_{5}}_{b_1...b_{4} d}
- \frac{9!}{2}\, 5! \,  \delta^{d_1...d_{5}}_{b_1...b_{4} c} \omega^{b_4...b_1
\hspace{2mm}c}_{\hspace{8mm}d} = 0\hspace{2mm}.
\ee
Now, contracting $d_5$ and $d$ in  \eqref{eq4} we get
\begin{equation}
\omega^{b_4...b_1  \hspace{2mm}c}_{\hspace{8mm}c}  =
- \frac{56}{9!} \nu \, (dA)^{b_1...b_4}  \hspace{2mm}
\end{equation}
and, inserting this in \eqref{eq4}, we find
\begin{equation}  \label{eq5}
\omega^{\left[d_1...d_4 \hspace{2mm} d_5   \right]}_{\hspace{9,5mm}d}
= -\nu \frac{4\cdot 2}{9!} \, (dA)^{[d_1...d_4} \delta^{\ d_5]}_d  \hspace{2mm}.
\end{equation}
We now use this equation to find  $\omega^{d_1...d_4 \hspace{2mm} d_5}_{\hspace{9mm}d} $ without
antisymmetrization. To this end, we use the following trick: first we make eq~\eqref{eq5}
more explicit, with $d_5$ interchanged with $d$, so that the antisymmetrization involves
$d_1,\cdots d_4$ and $d$,
\begin{eqnarray}
\nonumber    \omega^{d_1 d_2 d_3 d_4 d_5}_{\hspace{15mm} d} &-&
\omega^{d_1 d_2 d_3 \hspace{3mm} d_5 d_4}_{\hspace{10mm} d} \\
\nonumber
-  \omega^{d_1 d_2  \hspace{3mm} d_5 d_4 d_3}_{\hspace{7mm} d} &-& \omega^{d_1
\hspace{3mm} d_3 d_4 d_5 d_2}_{\hspace{4mm} d}\\
 - \omega^{\hspace{2mm}d_1 d_2 d_3 d_4 d_5}_{d} &=&- \frac{4\cdot 2}{9!} \nu
 \left( (dA)^{d_1 d_2 d_3 d_4 } \delta^{ d_5}_d \right. \\
\nonumber   - (dA)_{d}{}^{d_2d_3d_4} \eta^{d_5 d_1 }  &-&
(dA)^{d_1}{}_d{}^{d_3d_4}\eta^{d_5 d_2 } \\
\nonumber  - (dA)^{d_1d_2}{}_d{}^{d_4} \eta^{d_5 d_3 }   &-&
(dA)^{ d_1 d_2 d_3}{}_d \eta^{d_5 d_4 }\left. \right)\hspace{2mm}.
\end{eqnarray}
Antisymmetrizing the indices $d_1...d_5$ with weight one leads to
\be
\label{eq6}
\omega^{d_1...d_5}_{\hspace{8,5mm}d} - 4\cdot \omega^{\left[d_1...d_4 \hspace{2mm} d_5   \right]}_{\hspace{9,5mm}d} = -
\frac{4\cdot 2}{9!} \, \nu \, (dA)^{[d_1 d_2 d_3 d_4 } \delta^{ d_5]}_d  \hspace{2mm},
\ee
and using \eqref{eq5} in \eqref{eq6}, we finally obtain
\be \label{c1again}
\omega^{d_1...d_5}_{\hspace{8,5mm}d} = - \frac{40}{9!} \, \nu \, (dA)^{\left[d_1 ... d_4\right. }
\delta^{ d_5 \left. \right] }_d    \hspace{2mm}.
\ee
or, equivalently, \eqref{eq7}.

\subsection{Calculation of the terms in \eqref{riccieq} }
\label{a5}
Defining the zero-form matrix $\widehat{dA}= (dA)_{a_1\dots a_4}\gamma^{a_1\dots a_4}$, eq~\eqref{eq7}
may be rewritten as
\begin{equation}
\label{dAk}
\omega_5= -\frac{20}{9!} (\widehat{dA}e+e\widehat{dA})
\end{equation}
Inserting this relation into \eqref{riccieq}, we obtain
\begin{eqnarray}
\label{riccieqbis1}
30 Tr(  R_L e^9 )  &=&  48 \frac{20}{9!} \nu^2 dA \,   Tr(\widehat{dA} e^7 ) \\ \nonumber
&-&  54 \nu^2 \left( \frac{20}{9!}\right)^2  Tr(4 \widehat{dA}\widehat{dA}e^{11} + 3\widehat{dA}e\widehat{dA}e^{10}
+ 4 \widehat{dA}e^2\widehat{dA}e^{9}\\ \nonumber &+& 4\widehat{dA}e^3\widehat{dA}e^{8} +
4\widehat{dA}e^4\widehat{dA}e^{7} + 4 \widehat{dA}e^5\widehat{dA}e^{6})    \hspace{2mm}.
\end{eqnarray}
Let us now compute the terms in this equation. First, the trace on the $l.h.s.$ is given by
\begin{eqnarray}
\label{trRL}
Tr(  R_L e^9 )&=& \frac{1}{4} Tr( R_L^{b_1b_2}{}_{a_1a_2}\gamma_{b_1b_2} e^{a_1} e^{a_2} e^{a_3}
\dots e^{a_{11}}\gamma_{a_3\dots a_{11}}) \nonumber \\
&=& \frac{1}{4} Tr(\gamma_{b_1b_2}\gamma_{a_3\dots a_{11}}) R_L^{b_1b_2}{}_{a_1a_2} \epsilon^{a_1\dots a_{11}} E \nonumber \\
&=& 8 \epsilon_{b_1b_2a_3\dots a_{11}} \epsilon^{a_1\dots a_{11}} R_L^{b_1b_2}{}_{a_1a_2} E \nonumber \\
&=& -8 . 9! \delta^{a_1a_2}_{b_1b_2} R_L^{b_1b_2}{}_{a_1a_2} E \nonumber\\
&=& -16 . 9! R_L E \ ,
\end{eqnarray}
where $E$ is an $11$-form defined by $e^{a_1}\dots e^{a_{11}} = \epsilon^{a_1\dots a_{11}} E$,
and we have written $R^{b_1b_2}=R^{b_1b_2}{}_{a_1a_2} e^{a_1} e^{a_2}$.
The first term on the r.h.s. of \eqref{trRL} contains the form
\begin{eqnarray}
\label{ftotr}
\nonumber
   dA \, Tr( \widehat{dA} e^7 ) &=& 32  (dA)_{b_1...b_4}e^{b_1}...e^{b_4}
\, \epsilon_{a_1 ... a_{11}} (dA)^{a_1 ... a_4} e^{a_5}...e^{a_{11}} E\\
\nonumber &=&  32\, (dA)_{b_1...b_4}(dA)^{a_1...a_4}
\epsilon_{a_1 ... a_{11}}\epsilon^{b_1 ... b_4 a_5 ... a_{11}} E \\
\nonumber &=&  - 7!\cdot 32 \, (dA)_{b_1...b_4}\,
(dA)^{a_1...a_4} \,   \delta_{a_1 ... a_4}^{b_1 ... b_4} \, E \\
&=& -7! \cdot 4! \cdot 32  \,
(dA)_{a_1...a_4}\, (dA)^{a_1...a_4} E \hspace{2mm} ,
\end{eqnarray}
where, as before, we have written $dA= (dA)_{b_1...b_4}e^{b_1}...e^{b_4}$.

The calculation of the remaining terms is slightly more complicated. These
terms have the form
\begin{eqnarray}
Tr(\widehat{dA}e^k\widehat{dA}e^{11-k}) &=&
 Tr(\widehat{dA} \gamma^{a_1\dots a_k}\widehat{dA} \gamma^{a_{k+1}...a_{11}} \,
\epsilon_{a_1 ... a_{11}}E) \\
 \nonumber&=&  \,\frac{(-1)^{\frac{k(k-1)}{2}}}{k!}
 Tr(\widehat{dA}\gamma^{a_1\dots a_k}\widehat{dA}
 \gamma_{b_1\dots b_k})\epsilon^{b_1\dots b_k a_{k+1}...a_{11} }\epsilon_{a_1 ... a_{11}}\, E \\
 \nonumber &=& - (-1)^{\frac{k(k-1)}{2}} (11-k)! Tr(\widehat{dA} \gamma^{a_1\dots a_k}
 \widehat{dA} \gamma_{a_1\dots a_k})E \\ \nonumber
 &=& - 32 (-1)^{\frac{k(k-1)}{2}} 4!  (11-k)! N_k (dA)_{a_1...a_4}\, (dA)^{a_1...a_4} E \quad,
\end{eqnarray}
where we have used the property \eqref{geg}
and the numbers $N_k$ in the equation are defined through
\begin{equation}\label{Nk}
\gamma^{a_1\dots a_k}
 \widehat{dA} \gamma_{a_1\dots a_k} = N_k  \widehat{dA}\ .
\end{equation}
These numbers may be computed using gamma matrix algebra; alternatively, they can be found in Ref.
\cite{Pr:99}.
Their values are: $N_0=1$, $N_1=3$, $N_2=2$, $N_3=66$, $N_4=-144$, $N_5=1680$. Then, the second trace on the
r.h.s. of \eqref{riccieqbis1} is given by $-32\cdot 168\cdot 9!\cdot 4! (dA)^{a_1\dots a_4}
(dA)_{a_1\dots a_4}\, E$. When this is taken into account, eq. \eqref{riccieqbis1} reads
\begin{equation}
\label{RoFF}
R_L= 16\cdot \frac{7!\cdot 4!}{(9!)^2} \gamma^2 (dA)^{a_1\dots a_4} (dA)_{a_1\dots a_4}\ .
\end{equation}


\end{document}